\documentclass[prc,aps,floatfix,nofootinbib,twocolumn,superscriptaddress]{revtex4-2} 

\usepackage{graphicx}
\usepackage{color}
\usepackage{enumerate}
\usepackage{amssymb}
\usepackage{amsmath}
\usepackage{amsfonts}
\usepackage{bbm}
\usepackage{dsfont}
\usepackage{url}
\usepackage[colorlinks=true,allcolors=blue,breaklinks]{hyperref}

\renewcommand{\vec}[1]{\mbox{\boldmath $#1$}}

\begin{document}

\title{
Emulating multi-channel scattering based on the eigenvector continuation in the discrete basis formalism}

\author{K. Hagino}
\affiliation{ 
Department of Physics, Kyoto University, Kyoto 606-8502,  Japan} 

\author{Zehong Liao}
\affiliation{ 
Sino-French Institute of Nuclear Engineering and Technology, Sun Yat-sen University, Zhuhai 519082, China}
\affiliation{ 
Department of Physics, Kyoto University, Kyoto 606-8502,  Japan} 

\author{S. Yoshida}
\affiliation{ 
School of Data Science and Management, Utsunomiya University, Mine, Utsunomiya 321-8505, Japan
}
\affiliation{ 
RIKEN Nishina Center for Accelerator-based Science, RIKEN, Wako 351-0198, Japan
}

\author{M. Kimura}
\affiliation{ 
RIKEN Nishina Center for Accelerator-based Science, RIKEN, Wako 351-0198, Japan
}

\author{K. Uzawa}
\thanks{Present address: Nuclear Data Center, Japan Atomic Energy Agency, Tokai, Ibaraki 319-1195, Japan.} 
\affiliation{ 
Department of Physics, Kyoto University, Kyoto 606-8502,  Japan} 

\begin{abstract}
We construct an emulator for a multi-channel 
scattering problem based on the eigenvector continuation. 
To this end, we employ the Kohn variational principle formulated in the discrete basis formalism. 
We apply this to one-dimensional scattering problems with a Gaussian barrier. 
Both for a single-channel and two-channel problems, we demonstrate that the penetration probability 
as well as the wave functions are well reproduced by the emulator. In particular,  
the energy dependence of the penetrability in a wider range of energy from well below the barrier to 
well above the barrier is successfully reproduced. 
\end{abstract}

\maketitle

\section{Introduction}

An emulator provides a convenient tool to surrogate a quantum mechanical calculation. 
With an emulator, one can obtain an approximate solution, which however is highly accurate and quick to 
obtain. This significantly reduces a computational cost, providing a useful means when similar calculations 
have to be repeated many times. A typical application of emulators is to an uncertainty quantification~\cite{Dobaczewski2014,Boehnlein2022}, in 
which one has to repeat calculations many times with different parameter sets to estimate 
uncertainties of a theoretical model. 

The eigenvector continuation (EC) \cite{Frame2018,Sarkar2021,Duguet2024} has been a powerful 
tool to construct emulators. 
In this method, 
to obtain an approximate solution for a model Hamiltonian, 
one linearly superposes a few eigenvectors of the model Hamiltonian with several different 
model parameters. This method was firstly applied to bound state  
problems of atomic nuclei \cite{Ekstrom2019,Konig2020,Wesolowski2021,Luo2024,Franzke2024,Bonilla2022,Yoshida2022,Djarv2022,Demol2020,Yapa2023}. 
Recently, the eigenvector continuation has been applied also to nuclear reactions as well \cite{Furnstahl2020,Drischler2021,Melendez2021, Bai2021, Bai2022, Zhang2022,Liu2024,Odell2024,Garcia2023}. 

In this paper, we construct an emulator for nuclear scattering based on the eigenvector continuation 
together with the Kohn variational principle \cite{Kohn1948,Kamimura1977,Mito1976,Matsuse1978,Beck1975,Beck1975-2,Sharaf2024}. 
Even though the Kohn variational principle has been 
employed in Refs. \cite{Furnstahl2020,Garcia2023} to construct emulators for nuclear scattering, 
we reformulate it with a different formalism, that is, the discrete basis formalism \cite{Fanto2018,Bertsch2019,Alhassid2020,Alhassid2021}. 
In this formalism, a scattering problem is solved in a matrix form with a linear algebra. 
This has been applied to induced fission 
reactions \cite{Fanto2018,Bertsch2019,Alhassid2020,Alhassid2021} as well as to barrier penetration 
problems \cite{Hagino2024,Bertsch2024}. An advantage of this method is that it 
is well compatible with many-body wave functions \cite{Hagino2024-2}. 
Recently, a new scattering theory based on the discrete basis formalism with the Kohn variational 
principle has been proposed in Ref. \cite{Hagino2024-2}. The aim of this paper is applied it to construct 
scattering emulators using the eigenvector continuation. 

The paper is organized as follows. 
In Sec.~\ref{sec:SingleChannel}, we consider a single channel scattering with a Gaussian barrier. 
We first detail the discrete basis formalism for a one-dimensional scattering, and then 
construct an emulator based on the eigenvector continuation. 
In Sec.~\ref{sec:TwoChannel}, we extend the scattering emulator to a multi-channel problem. 
We then summarize the paper in Sec.~\ref{sec:Summary}.

\section{Single-channel Problem} \label{sec:SingleChannel}

\subsection{Discrete basis formalism}

We consider a one-dimensional problem with a potential $V(x;\vec{\theta})$, where 
$\vec{\theta}$ is a set of parameters which characterize the potential $V$. 
In this paper, we consider a Gaussian potential given by, 
\begin{equation}
V(x;\vec{\theta})=V_0\,e^{-x^2/2s^2}, 
\label{eq:Gauss}
\end{equation}
in which the parameter set $\vec{\theta}$ consists of the height $V_0$ and the width $s$ of the 
barrier, that is, $\vec{\theta}=(V_0,s)$. 
The Hamiltonian of this system is given by
\begin{equation}
H=-\frac{\hbar^2}{2m}\,\frac{d^2}{dx^2}+V(x;\vec{\theta}), 
\label{eq:Hsingle}
\end{equation}
where $m$ is the mass of a particle. 
We consider a situation in which the particle is incident from the left hand side of the barrier. 
The boundary condition then reads,
\begin{eqnarray}
    \phi(x)&\to& e^{ikx}+Re^{-ikx}~~~(x\to-\infty), 
    \label{eq:bc1}
    \\
    &\to&Te^{ikx}~~~~~~~(x\to\infty), 
    \label{eq:bc2}
\end{eqnarray}
where $\phi(x)$ is the wave function and $k$ is the wave number. $T$ and $R$ are the transmission and the 
reflection coefficients, respectively. 
The penetration probability of the barrier is given by, 
\begin{equation}
P=|T|^2=1-|R|^2,
\label{eq:P}
\end{equation}
in which we have used the flux conservation. 

Let us solve this problem with the Kohn variational principle in the discrete basis 
formalism \cite{Hagino2024-2}. For this purpose, we first discretize the coordinate $x$ 
with $N$ mesh points from 
$x_{\rm min}$ to $x_{\rm max}$. 
That is, 
the $i$-th mesh point is given by 
$x_i=x_{\rm min}+(i-1)\Delta x$ with $i=1,2,\cdots, N$, where $\Delta x=(x_{\rm max}-x_{\rm min})/(N-1)$ is 
the mesh spacing. 
Here, the potential $V$ is assumed to be vanishingly small at $x_{\rm min}$ and $x_{\rm max}$ so that the wave 
functions at these points satisfy the asymptotic boundary conditions, Eq.~\eqref{eq:bc1} and Eq.~\eqref{eq:bc2}. 
Using the three-point formula for the second derivative, the 
Hamiltonian, Eq.~\eqref{eq:Hsingle}, is represented in a matrix form as,
\begin{equation}
H_{ij}=-t\,\delta_{i,j+1}+(2t+V_i)\,\delta_{i,j}-t\,\delta_{i,j-1},
\end{equation}
with $i,j=1,2,\cdots,N$, 
where $t\equiv\hbar^2/[2m(\Delta x)^2]$ and $V_i\equiv V(x_i;\vec{\theta})$. 
The wave function $\phi_i\equiv\phi(x_i)$ then obeys the equation
\begin{equation}
    -t\phi_0\,\delta_{i,1}+\sum_{j=1}^NH_{ij}\phi_j-t\phi_{N+1}\delta_{i,N}=E\phi_i,
    \label{eq:Seq}
\end{equation}
where $E$ is the energy of the system. 
In addition, the wave function $\phi_0$ at $x=x_0$ satisfies the equation
\begin{equation}
    -t\phi_{-1}+2t\phi_0-t\phi_1=E\phi_0. 
    \label{eq:Seq0}
\end{equation}
Notice that the wave function $\phi_{i}^{(0)}$ for a free particle is given by 
\begin{equation}
-t(\phi_{i+1}^{(0)}-2\phi_{i}^{(0)}+\phi_{i-1}^{(0)})=E\phi_{i}^{(0)}.
\end{equation}
Substituting $\phi_{i}^{(0)}\propto e^{\pm ikx_i}$ into this equation, 
one obtains the dispersion relation 
\begin{equation}
    \cos(k\Delta x)=1-\frac{E}{2t}. 
    \label{eq:dispersion}
\end{equation}

\begin{figure}[tb] 
\begin{center} 
\includegraphics[width=0.9\columnwidth]{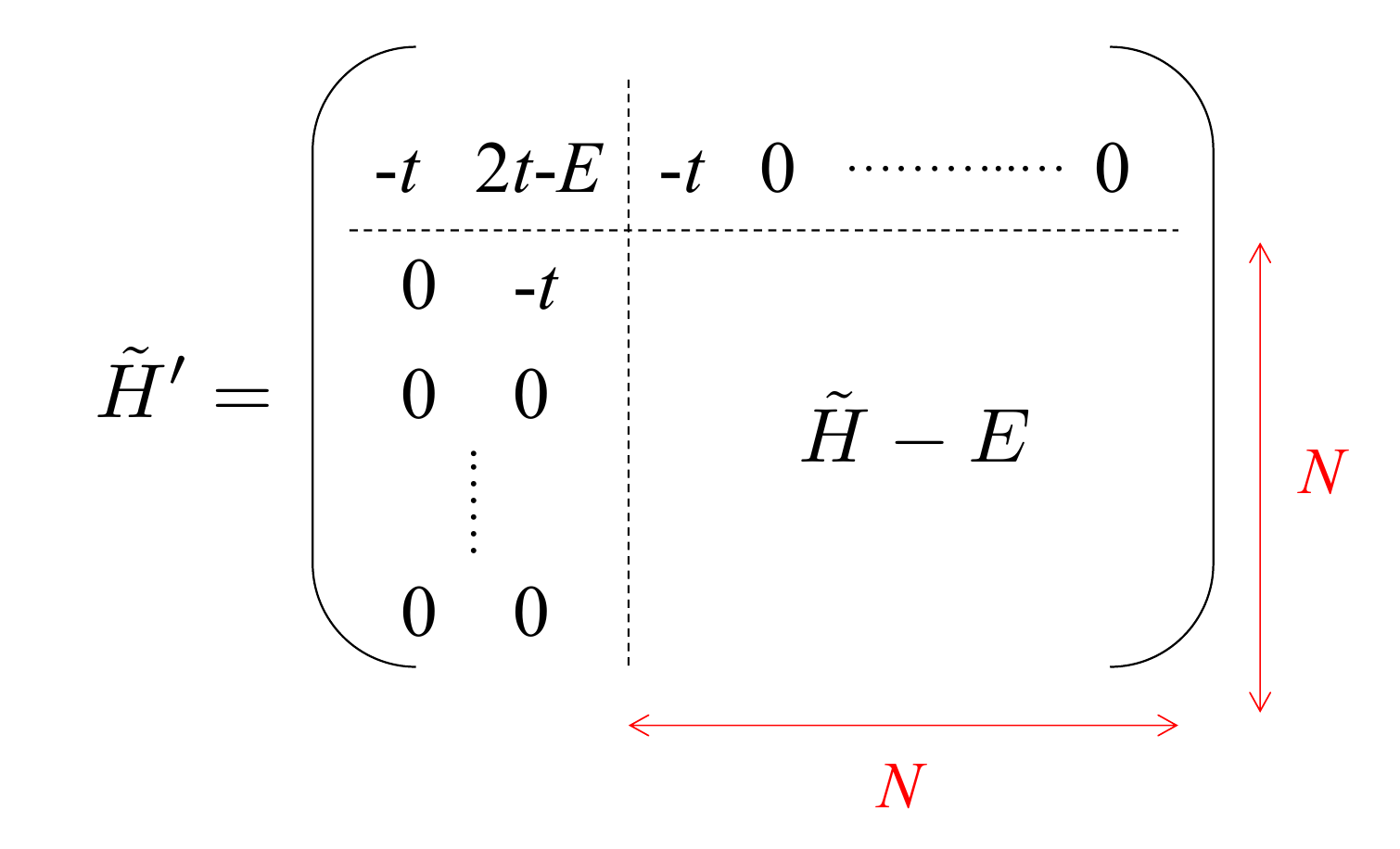} 
\caption{
A schematic illustration of the 
the $(N+1)\times(N+2)$ matrix $\tilde{H}'_{ij}$ defined by Eqs.~\eqref{eq:Hdashed1}--\eqref{eq:Hdashed2}.
}
\label{fig:t}
\end{center} 
\end{figure} 

The last term on the left hand side of Eq.~\eqref{eq:Seq} can be handled by noticing 
$\phi_{N+1}=e^{ik\Delta x}\,\phi_N$ (see Eq.~\eqref{eq:bc2}). 
This is equivalent to modifying Eq.~\eqref{eq:Seq} to, 
\begin{equation}
    -t\phi_0\,\delta_{i,1}+\sum_{j=1}^N\tilde{H}_{ij}\phi_j=E\phi_i,
    \label{eq:Seq2}
\end{equation}
where $\tilde{H}_{ij}\equiv H_{ij}$ except for $\tilde{H}_{NN}\equiv H_{NN}-te^{ik\Delta x}$. 
The $N+1$ equations, Eq.~\eqref{eq:Seq0} and Eq.~\eqref{eq:Seq2}, can be combined into 
\begin{equation}
\tilde{H}'
 \left(\begin{matrix}
\phi_{-1} \\
\phi_0 \\
\phi_1 \\
\vdots \\
\phi_N
\end{matrix}\right)=0, 
\label{eq:Hphi}
\end{equation}
where 
the $(N+1)\times(N+2)$ matrix $\tilde{H}'_{ij}$ with $i=0,1,\cdots,N$ and $j=-1,0,1,\cdots,N$ is defined as (see Fig. 1), 
\begin{eqnarray}
\tilde{H}'_{ij}&=&\tilde{H}_{i,j}-E\,\delta_{i,j},~~~(i,j\geq 1), \label{eq:Hdashed1} \\
\tilde{H}'_{0,-1}&=&\tilde{H}'_{01}=\tilde{H}'_{10}=-t, \\
\tilde{H}'_{00}&=&2t-E, \\
\tilde{H}'_{1,-1}&=&0.  \label{eq:Hdashed2}
\end{eqnarray}

The wave function $\phi_0$ at $x=x_0$ in the first term on the left hand side of 
Eq.~\eqref{eq:Seq2} has both the incoming and the outgoing components (see Eq.~\eqref{eq:bc1}). 
In order to take into account this term, we employ the Kohn variational principle. 
For that purpose, we extend the model space by adding $x_{-1}$ and $x_0$, and introduce the 
basis functions 
\begin{eqnarray}
\psi_k&=&(e^{-ik\Delta x},1,0,\cdots,0)^T, \\
\psi_{-k}&=&(\psi_k)^*=(e^{ik\Delta x},1,0,\cdots,0)^T, \\
\psi_n&=&(0,0,0,\cdots,1,0,\cdots,0)^T,
\end{eqnarray}
where $T$ represents transpose. Each of these basis functions has $N+2$ components, 
for which $\psi_n$ with $n=1,2,\cdots,N$ has a component of $\psi_n(i)=\delta_{i,n+2}$. 
We express the total wave function as 
\begin{equation}
    \Psi=\psi_k+b_0\psi_{-k}+\sum_{n=1}^Nb_n\psi_n\equiv \psi_k+\sum_{n=0}^Nb_n\psi_n, 
    \label{eq:wf_tot}
\end{equation}
and impose the condition $\tilde{H}'\Psi=0$. Here, $\psi_0$ is defined as $\psi_0\equiv\psi_{-k}$. 
Defining the $(N+2)\times(N+1)$ wave function matrix $\varphi_{ij}\equiv \psi_j(i)$ with $i=-1,0,1,\cdots,N$ and $j=0,1,\cdots,N$, 
the condition $\tilde{H}'\Psi=0$ reads 
\begin{equation}
\sum_{i=-1}^{N}\sum_{j=0}^{N}\tilde{H}'_{li}\varphi_{ij}b_j=-\sum_{i=-1}^{N}\tilde{H}'_{li}\psi_k(i)~~~~(l=0,1,\cdots,N),
\end{equation}
or
\begin{equation}
\sum_{j=0}^{N}(\tilde{H}'\varphi)_{lj}b_j=-(\tilde{H}'\psi_k)_l\equiv -\tilde{\psi}_l~~~(l=0,1,\cdots,N). 
\end{equation}
The amplitudes $\{b_i\}$ can then be obtained by inverting the $(N+1)\times (N+1)$ matrix $(\tilde{H}'\varphi)$ as 
\begin{equation}
b_j=-\sum_{l=0}^{N}\left[(\tilde{H}'\varphi)^{-1}\right]_{jl}\tilde{\psi}_l.
\label{eq:original}
\end{equation}
The penetrability (\ref{eq:P}) is then computed as,
\begin{equation}
    P=1-|b_0|^2=|b_N|^2.
\end{equation}

\subsection{Eigenvector continuation}

Using the solution obtained in the previous subsection, one can 
construct a vector $\phi$ defined as
\begin{equation}
    \phi=\sum_{n=1}^Nb_n\psi_n=(0,0,b_1,b_2,\cdots,b_N)^T.
\end{equation}
The idea of the eigenvector continuation is to linearly superpose in Eq.~\eqref{eq:wf_tot} 
\begin{equation}
    \phi_i=(0,0,b^{(i)}_1,b^{(i)}_2,\cdots,b^{(i)}_N)^T, 
\end{equation}
obtained with the parameter set $\vec{\theta}_i$ instead of $\psi_n$ 
and obtain the scattering solution for the parameter set $\vec{\theta}$. That is,
\begin{equation}
    \Psi=\psi_k+c_0\psi_{-k}+\sum_{i=1}^{N_{\rm EC}}c_i\phi_i,
\end{equation}
where $N_{\rm EC}$ is the number of the basis states to be superposed in the eigenvector continuation 
and $\{c_i\}$ are the coefficients for the superposition. 
From the condition of $\tilde{H}'\Psi=0$, one obtains 
\begin{equation}
\sum_{j=0}^{N_{\rm EC}}c_j\langle\phi'_i|\tilde{H}'|\phi_j\rangle
=-\langle\phi'_i|\tilde{H}'|\psi_k\rangle\equiv -d_i,
\label{eq:eqEC}
\end{equation}
where $\phi_0\equiv \psi_{-k}$ as in the previous subsection, and the function $\phi'_i$ is defined as 
$|\phi'_i\rangle\equiv \tilde{H}'|\phi_i\rangle$
\footnote{
We have examined another choice of $\phi'$ 
with 
$\phi'_i=(0,b^{(i)}_1,b^{(i)}_2,\cdots,b^{(i)}_N)^T$, 
that is, it is the same as $\phi_i$ but the first component is removed. 
We have found that this choice sometimes leads to an unstable solution, even though 
the flux conservation is better reproduced. 
}.
$d_i$ is defined as $d_i\equiv\langle\phi'_i|\tilde{H}'|\psi_k\rangle$. 
The matrix elements on the left hand side of Eq.~\eqref{eq:eqEC} can 
be easily obtained by noticing that 
$\Psi_i=\psi_k+b_0^{(i)}\psi_{-k}+\phi_i$ satisfies 
$\tilde{H}_i\Psi_i=0$, where $\tilde{H}_i$ is the matrix 
$\tilde{H}$ constructed with the potential 
$V(x;\vec{\theta}_i)$ and that 
$\tilde{H}_i-\tilde{H}$ contains only the diagonal 
matrix $[(V(x;\vec{\theta}_i)-V(x;\vec{\theta})]\,\vec{1}$, 
where $\vec{1}$ is the unit matrix. 
The coefficients $c_i$ in Eq.~\eqref{eq:eqEC} can be obtained 
by inverting the matrix $A_{ij}\equiv\langle\phi'_i|\tilde{H}'|\phi_j\rangle $ as,
\begin{equation}
c_i=\sum_{j=0}^{N_{\rm EC}}(A^{-1})_{ij}d_j.
\label{eq:ec}
\end{equation}
The penetrability is then computed as
\begin{equation}
    P=1-|c_0|^2=\left|\sum_{i=1}^{N_{\rm EC}}c_ib_N^{(i)}\right|^2.
\end{equation}

\begin{figure}[tb] 
\begin{center} 
\includegraphics[width=0.9\columnwidth]{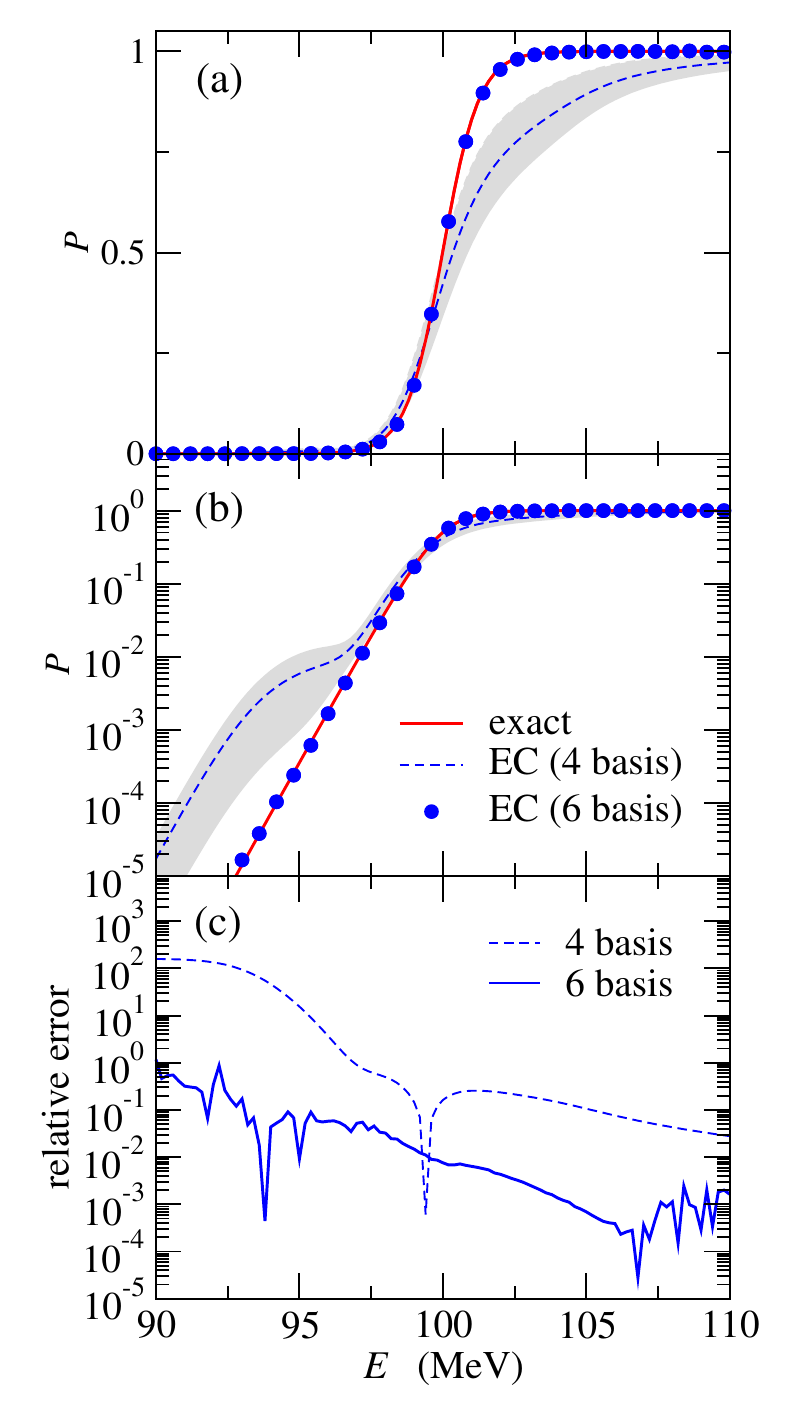} 
\caption{
(a) The penetrability of the Gaussian barrier with $V_0=100$ MeV 
in the linear scale as a function of the energy $E$. The solid 
line shows the exact result, while the dashed line 
and the dots are obtained 
with the eigenvector continuation with 
$N_{\rm EC}=4$ and $N_{\rm EC}=6$, respectively. 
For the eigenvector continuation, the values of $V_0$ 
are chosen randomly in the range of 
(95 MeV,105 MeV) and 
an ensemble average is taken for 5 different samples. 
For $N_{\rm EC}=4$, the standard errors are also shown 
by the gray area, 
while the errors are within the symbols for $N_{\rm EC}=6$. 
(b) The same as (a) but in the logarithmic scale. 
(c) The relative error defined by 
$|P_{\rm exact}-P_{\rm EC}|/P_{\rm exact}$, where $P_{\rm exact}$ and $P_{\rm EC}$ are the exact 
penetrability and that obtained with the eigenvector 
continuation, respectively. 
The dashed and the solid lines denote the results 
for $N_{\rm EC}=4$ and $N_{\rm EC}=6$, respectively. 
}
\end{center} 
\end{figure} 

Fig. 2(a) and 2(b) show the penetrability for the Gaussian 
barrier (\ref{eq:Gauss}) as a function of $E$ 
in the linear and the logarithmic scales, respectively. 
To this end, we use 
$V_0=100$ MeV, $s=3$ fm, 
and $m=29 m_N$, $m_N$ being the nucleon mass, to mimic 
the fusion reaction of $^{58}$Ni+$^{58}$Ni 
\cite{Hagino2024,Dasso1983,Dasso1983-2,Hagino2004}. 
We discretize the coordinate $x$ from $x_{\rm min}=-15$ fm to 
$x_{\rm max}$=15 fm 
with $\Delta x=$ 0.05 fm. 
In the figure, the solid lines show the exact solution, while 
the dashed lines and the dots 
are obtained with the eigenvector continuation. 
For the latter, 
we take $N_{\rm EC}=4$ and 6 for the dashed lines and the dots, 
respectively, with $V_0$ randomly chosen in the range of 
(95 MeV,105 MeV), all with $s=3$ fm. 
The eigenvector continuation is applied at each $E$ separately. 
We take 5 samples of the training sets and take an average 
over the results with each sample. 
For $N_{\rm EC}=4$, we also show the standard errors. The errors 
are small enough for $N_{\rm EC}=6$ and are within the symbols in the figure. 
It is remarkable that the eigenvector continuation 
with $N_{\rm EC}=6$ reproduces 
the exact results not only around the barrier but also well below 
the barrier at which the penetrability is exponentially small. 
On the other hand, the errors are large for $N_{\rm EC}=4$ 
and the resultant energy dependence of penetrabilities are not consistent with the exact result.
We have confirmed that the results do not significantly change even if the number of samples 
is increased.
Notice that the dimension of the matrix to be inverted in Eq.~\eqref{eq:original} is $N+1$=602, 
which is significantly reduced to $N_{\rm EC}+1$=7 in Eq.~\eqref{eq:ec} in the eigenvector continuation 
with $N_{\rm EC}=6$. 
The computation time, measured by repeating the same 
calculations 1000 times, 
is reduced by a factor of about 10 with 
the eigenvector continuation
\footnote{The larger gain could be achieved 
under the condition that 
the inversion of the matrix dominates the computation time.}.
This factor remains almost 
the same, even when the number of mesh points 
is reduced to 302. 
It is interesting to see that the exact results are yet reproduced well even with such a small number 
of basis. 
Fig. 2(c) shows the relative error $|P_{\rm exact}-P_{\rm EC}|/P_{\rm exact}$, where $P_{\rm exact}$ and $P_{\rm EC}$ are the exact 
penetrability and that obtained with the eigenvector continuation, 
respectively. The dashed and the solid 
lines show the results with 
$N_{\rm EC}=4$ and $N_{\rm EC}=6$, respectively. 
One can see that the eigenvector continuation 
with $N_{\rm EC}=6$
reproduces the 
exact result within about $10^{-4}$ at energies above the barrier, 
even though the error increases to $O(10^{-1})$ at energies below 
the barrier. We also 
notice that the performance of the eigenvector 
continuation increases 
as $N_{\rm EC}$ is increased from 4 to 6.

\begin{figure}[tb] 
\begin{center} 
\includegraphics[width=0.9\columnwidth]{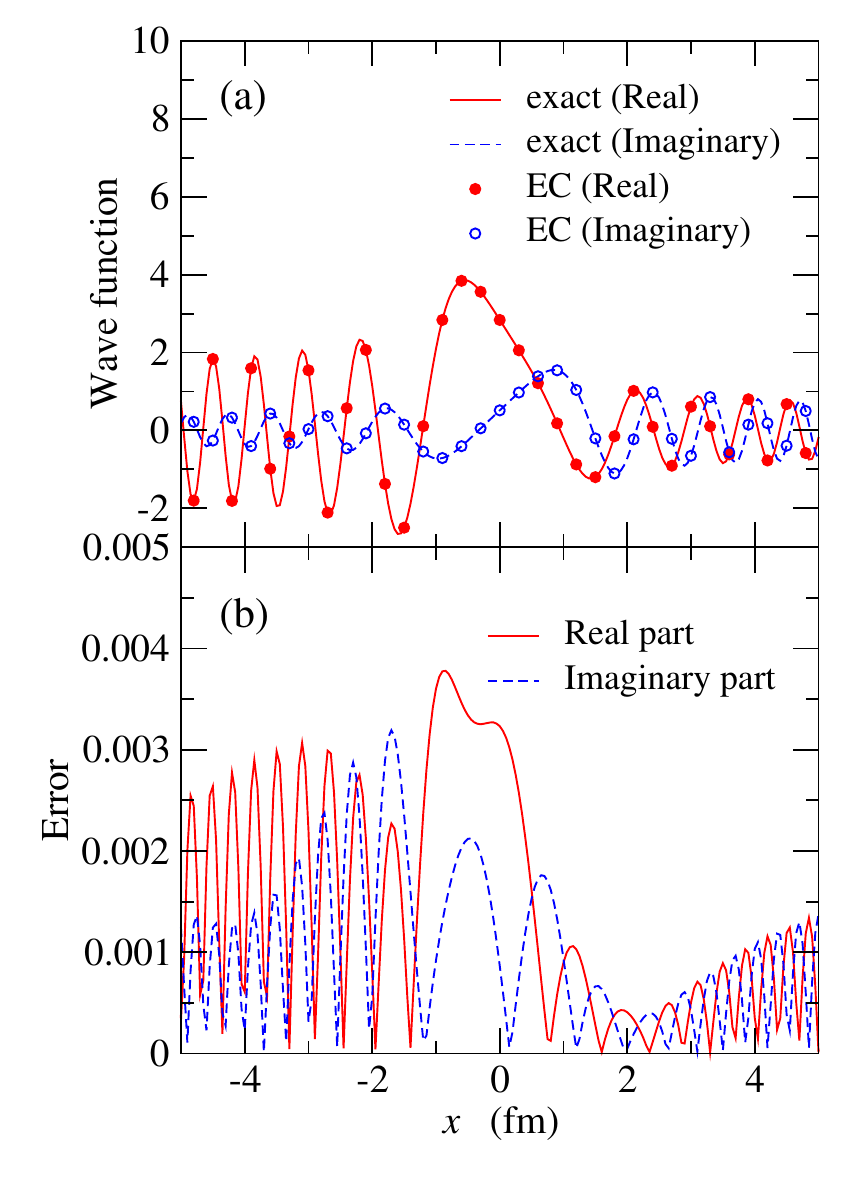} 
\caption{
(a) The wave function at the barrier top energy, $E=100$ MeV. 
The solid and the dashed lines show the real and the imaginary 
parts of the exact wave function, $\Psi_{\rm exact}$. 
The solid and the open circles show the real and the imaginary 
parts of the wave function obtained with the eigenvector 
continuation, $\Psi_{\rm EC}$. 
(b) The error of the wave function defined as 
$|\Psi_{\rm exact}-\Psi_{\rm EC}|$. 
}
\end{center} 
\end{figure} 

The upper panel of Fig. 3 shows the total wave function $\Psi$ 
at the barrier-top energy, $E=V_0=100$ MeV. 
The solid and the dashed lines show the real and the imaginary 
parts of the exact wave function, and the solid and the open circles 
show the corresponding wave functions in the eigenvector 
continuation with $N_{\rm EC}=6$. 
The error of the wave function, 
$|\Psi_{\rm exact}-\Psi_{\rm EC}|$, is 
shown in the lower panel of Fig. 3. 
Here, $\Psi_{\rm exact}$ 
and $\Psi_{\rm EC}$ are the exact wave function and the wave 
function in the eigenvector continuation, respectively. Notice that 
we do not divide the error 
by $\Psi_{\rm exact}$, as it can be zero at the nodal points. 
One can see that the wave function is reproduced within $O(10^{-3})$ 
both for the real and the imaginary parts. This accuracy is sufficient 
to yield the relative error of $O(10^{-2})$ for the penetrability $P$, 
which requires only the asymptotic behavior of the wave function. 

\section{Two-channel problem} \label{sec:TwoChannel}

\subsection{Discrete basis formalism}

It is easy to extend the formalisms presented in the last section 
to a multi-channel problem. To demosntrate this, let us consider 
a two-channel problem given by, 
\begin{eqnarray}
&&-\frac{\hbar^2}{2m}\,\frac{d^2}{dx^2}
 \left(\begin{matrix}
\phi_{1}(x) \\
\phi_2(x) 
\end{matrix}\right) \nonumber \\
&&+ 
\left(\begin{matrix}
V(x;\vec{\theta}) & F(x;\vec{\theta}) \\
F(x;\vec{\theta}) & V(x;\vec{\theta})+\epsilon 
\end{matrix}
\right)
\left(\begin{matrix}
\phi_{1}(x) \\
\phi_2(x) 
\end{matrix}\right)
=
E
\left(\begin{matrix}
\phi_{1}(x) \\
\phi_2(x) 
\end{matrix}\right), 
\label{eq:cc}
\end{eqnarray}
where $\phi_1(x)$ and $\phi_2(x)$ are the scattering wave functions in the channels 1 and 2, respectively. 
$F(x;\vec{\theta})$ is the coupling potential and $\epsilon$ is the excitation energy of the channel 2. 
In this paper, 
we consider a Gaussian function for $F(x;\vec{\theta})$, 
\begin{equation}
F(x;\vec{\theta})=F_0\,e^{-x^2/2s_f^2}, 
\label{eq:Gauss2}
\end{equation}
together with the Gaussian potential for $V$ given by Eq.~\eqref{eq:Gauss}. 
In this case, a parameter set $\vec{\theta}$ consists of four variables,  $\vec{\theta}=(V_0,s,F_0,s_f)$.  
The scattering boundary conditions for the wave function $\phi_s(x)~(s=1,2)$ read,
\begin{eqnarray}
    \phi_s(x)&\to& e^{ik_sx}\delta_{s,1}+R_se^{-ik_sx}~~~(x\to-\infty), 
    \label{eq:bc1-2}
    \\
    &\to&T_se^{ik_sx}~~~~~~~(x\to\infty), 
    \label{eq:bc2-2}
\end{eqnarray}
where $k_s$ is the wave function for the channel $s$, given by 
$k_1=k$ in Eq.~\eqref{eq:dispersion} and 
\begin{equation}
    \cos(k_2\Delta x)=1-\frac{E-\epsilon}{2t}. 
\end{equation}
Notice that the incident right-going wave exists at $x=-\infty$ only for the channel 1.   
The penetration probability of the barrier is then given by, 
\begin{equation}
P=|T_1|^2+\frac{k_2}{k_1}|T_2|^2=1-|R_1|^2-\frac{k_2}{k_1}|R_2|^2.
\end{equation}

The discretized version of Eq.~\eqref{eq:cc} that corresponds to Eq.~\eqref{eq:Seq2} is given by, 

\begin{equation}
    -t\phi_{s,0}\,\delta_{i,1}+\sum_{s'=1,2}\sum_{j=1}^N\tilde{H}_{si,s'j}\phi_{s',j}=E\phi_{s,i}, 
\end{equation}
where 
$\phi_{s,i}$ is defined as $\phi_{s}(x_i)$ and $\tilde{H}_{si,s'j}$ is given by,
\begin{eqnarray}
\tilde{H}_{si,s'j}&=&\left[-t\,\delta_{i,j+1}+(2t+V_i+\epsilon\,\delta_{s,2})\,\delta_{i,j}-t\,\delta_{i,j-1}\right]\delta_{s,s'} \nonumber \\
&&+F_i\delta_{i,j}\delta_{s\neq s'}
-te^{ik_s\Delta x}\delta_{s,s'}\delta_{i,j}\delta_{i,N}. 
\end{eqnarray}
Here $F_i$ is defined as $F_i\equiv F(x_i;\vec{\theta})$, and the last term 
is due to the relation $\phi_{s,N+1}=e^{ik_s\Delta x}\phi_{s,N}$ (see Eq.~\eqref{eq:bc2-2}). 
Equation (\ref{eq:Hphi}) can also be extended to the multi-channel system by introducing the channel indices as, 
\begin{equation}
\tilde{H}'
 \left(\begin{matrix}
\phi_{1,-1} \\
\phi_{2,-1} \\
\phi_{1,0} \\
\phi_{2,0} \\
\phi_{1,1} \\
\phi_{2,1} \\
\vdots \\
\phi_{1,N} \\
\phi_{2,N}
\end{matrix}\right)=0, 
\end{equation}
with 
\begin{eqnarray}
\tilde{H}'_{si,s'j}&=&\tilde{H}_{si,s'j}-E\,\delta_{i,j}\delta_{s,s'},~~~(i,j\geq 1), \\
\tilde{H}'_{s0,s'-1}&=&\tilde{H}'_{s0,s'1}=\tilde{H}'_{s1,s'0}=-t\,\delta_{s,s'}, \\
\tilde{H}'_{s0,s'0}&=&(2t-E+\epsilon\,\delta_{s,2})\,\delta_{s,s'}, \\
\tilde{H}'_{s1,s'-1}&=&0.  
\end{eqnarray}
Notice that the dimension of $\tilde{H}'$ is 
$2(N+1)\times 2(N+2)$. 

The basis functions for the Kohn variational principle can also be extended in a similar way. 
By arranging the wave functions as
$\psi=(\phi_{1,-1},\phi_{2,-1},\phi_{1,0},\phi_{2,0},\cdots,\phi_{1,N},\phi_{2,N})^T$, 
the basis functions are given by 
\begin{eqnarray}
\psi_k&=&(e^{-ik\Delta x},0,1,0,\cdots,0)^T, \\
\psi^{(1)}_{-k}&\equiv&\psi^{(1)}_0=(e^{ik_1\Delta x},0,1,0,\cdots,0)^T, \\
\psi^{(2)}_{-k}&\equiv&\psi^{(2)}_0=(0,e^{ik_2\Delta x},0,1,0,\cdots,0)^T, \\
\psi^{(1)}_n(i)&=&\delta_{i,4+2n-1}, \\
\psi^{(2)}_n(i)&=&\delta_{i,4+2n}.
\end{eqnarray}
With these basis functions, the total wave function is given by
\begin{equation}
    \Psi=\psi_k+\sum_{s=1,2}\sum_{n=0}^Nb_{sn}\psi^{(s)}_n, 
\end{equation}
with which it is guaranteed that the incident right-going wave exists only for the channel 1.  
Defining the $2(N+2)\times 2(N+1)$ wave function matrix $\varphi_{i,sj}\equiv \psi^{(s)}_j(i)$, 
the condition $\tilde{H}'\Psi=0$ is transformed to 
\begin{equation}
\sum_{s'=1,2}\sum_{j=0}^{N}(\tilde{H}'\varphi)_{sl,s'j}b_{s'j}=
-(\tilde{H}'\psi_k)_{sl}\equiv -\tilde{\psi}_{sl},  
\end{equation}
from which the amplitudes $\{b_{si}\}$ can be obtained 
as
\begin{equation}
b_{sj}=-\sum_{s'=1,2}\sum_{l=0}^{N}\left[(\tilde{H}'\varphi)^{-1}\right]_{sj,s'l}\tilde{\psi}_{s'l}.
\end{equation}
The penetrability (\ref{eq:P}) is then computed as,
\begin{equation}
    P=1-\sum_{s=1,2}\frac{k_s}{k_1}|b_{s0}|^2=\sum_{s=1,2}\frac{k_s}{k_1}|b_{sN}|^2.
\end{equation}

\subsection{Eigenvector continuation\label{subsec:TwoChannEC}}

It is now straightforward to extend the eigenvector continuation for the single-channel problem 
to the multi-channel problem. 
To this end, we first construct a vector $\phi$ defined as
\begin{eqnarray}
    \phi&=&\sum_{s=1,2}\sum_{n=1}^Nb_{sn}\psi^{(s)}_n \\
    &=&(0,0,0,0,b_{11},b_{21},b_{12},b_{22},\cdots,b_{1N},b_{2N})^T.
\end{eqnarray}
We construct a similar function $\phi_i$ for each of the parameter set $\vec{\theta}_i$ of the potential 
and superpose them to construct the total wave function as 
\begin{equation}
    \Psi=\psi_k+c^{(1)}_0\psi^{(1)}_{-k}+c^{(2)}_0\psi^{(2)}_{-k}+\sum_{i=1}^{N_{\rm EC}}c_i\phi_i
    \equiv \psi_k+\sum_{i=-1}^{N_{\rm EC}}c_i\phi_i, 
\end{equation}
with $c_{-1}\equiv c^{(1)}_0$, $c_{0}\equiv c^{(2)}_0$, $\phi_{-1}\equiv \psi^{(1)}_{-k}$, 
and $\phi_{0}\equiv \psi^{(2)}_{-k}$. 
From the condition of $\tilde{H}'\Psi=0$, one obtains 
\begin{equation}
\sum_{j=-1}^{N_{\rm EC}}c_j\langle\phi'_i|\tilde{H}'|\phi_j\rangle
=-\langle\phi'_i|\tilde{H}'|\psi_k\rangle\equiv -d_i,
\end{equation}
where $\phi'_i$ and $d_i$ are defined as 
$\phi'_i\equiv\tilde{H}'|\phi_i\rangle$ and 
$d_i\equiv\langle\phi'_i|\tilde{H}'|\psi_k\rangle$, respectively, 
as in the 
single-channel problem. 
The coefficients $c_i$ can then be obtained as 
\begin{equation}
c_i=-\sum_{j=-1}^{N_{\rm EC}}(A^{-1})_{ij}d_j, 
\end{equation}
where 
$A_{ij}\equiv\langle\phi'_i|\tilde{H}'|\phi_j\rangle$. 
The penetrability is then computed as
\begin{equation}
    P=1-|c_{-1}|^2-\frac{k_2}{k_1}|c_0|^2=
    \left|\sum_{i=1}^{N_{\rm EC}}c_ib_{1N}^{(i)}\right|^2+\frac{k_2}{k_1}\left|\sum_{i=1}^{N_{\rm EC}}c_ib_{2N}^{(i)}\right|^2.
\end{equation}

\begin{figure}[tb] 
\begin{center} 
\includegraphics[width=0.9\columnwidth]{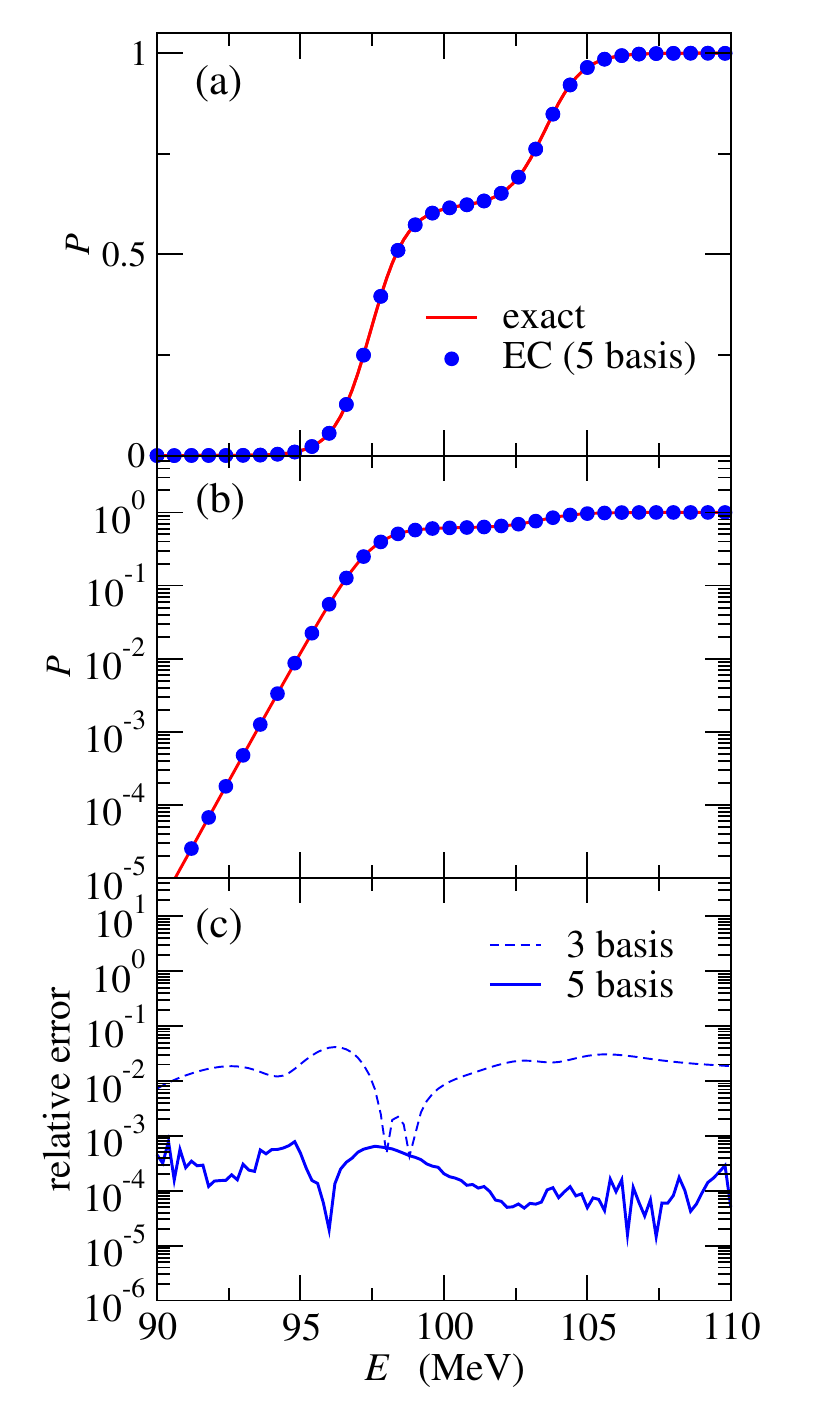} 
\caption{
Same as Fig. 2, but for the two-channel problem with the Gaussian potentials Eq.~\eqref{eq:Gauss} and Eq.~\eqref{eq:Gauss2}. The parameters of the potentials are 
$V_0=100$ MeV, $F_0=3$ MeV, and $s_f=3$ fm, and the excitation energy of the channel 2 
is set to be $\epsilon=$ 1 MeV. 
For the eigenvector continuation, the number of basis vectors are taken to be 
$N_{\rm EC}=5$, for which the strength of the coupling potential is varied randomly 
in the range of $F_0=$ (1 MeV, 5 MeV)  
while keeping the values of $V_0$, $s$, and $s_f$. 
}
\end{center} 
\end{figure} 

\begin{figure}[tb] 
\begin{center} 
\includegraphics[width=1.1\columnwidth]{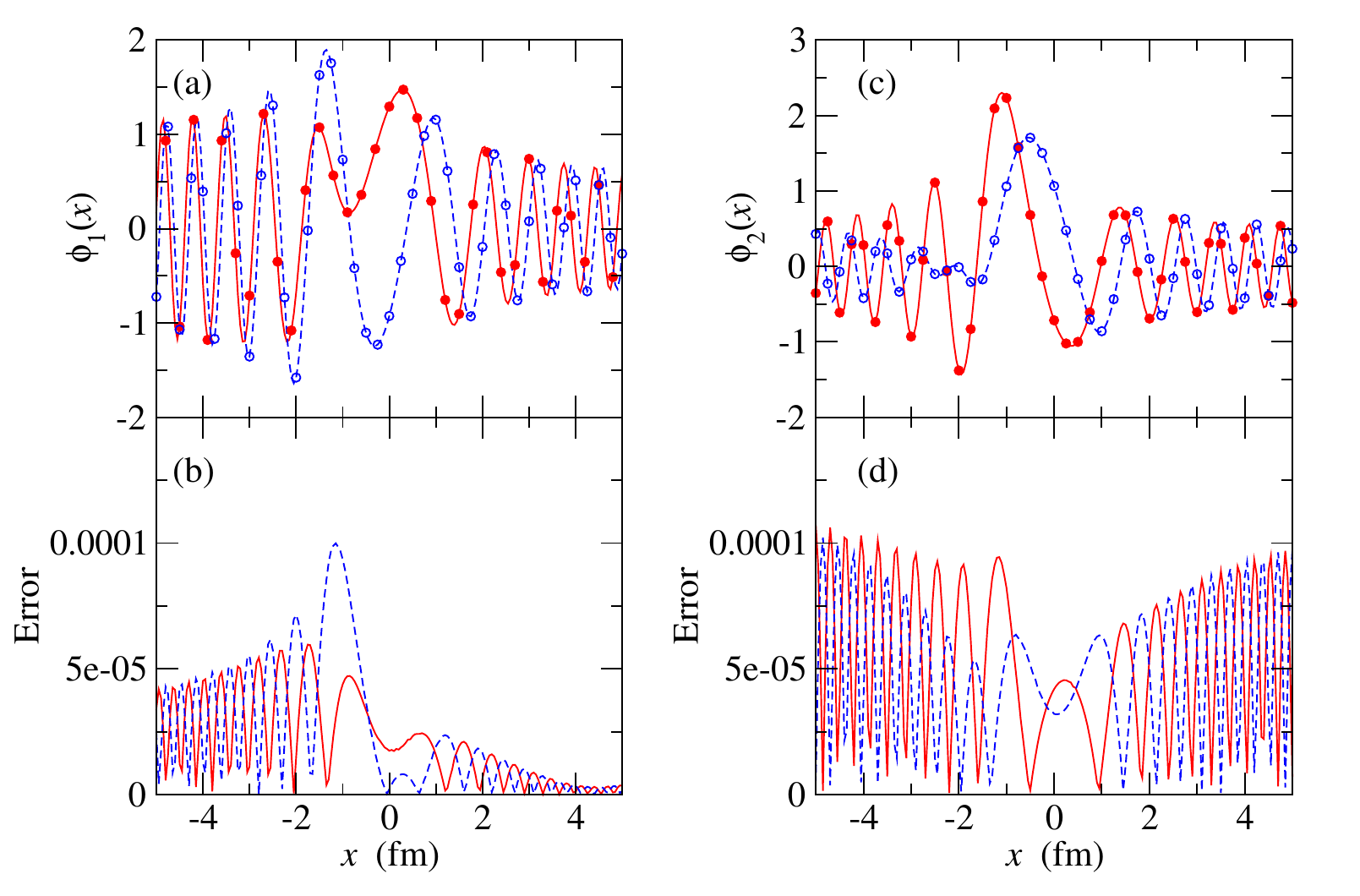} 
\caption{
Same as Fig. 3, but for the two-channel problem. 
See the caption of Fig. 4 for the parameters. 
The left and the right panels show the wave functions for the channel 1 and the channel 2, 
respectively. 
}
\end{center} 
\end{figure} 

Figure 4 shows the penetrability and the relative error as a function of energy $E$, for the Gaussian potentials, Eq.~\eqref{eq:Gauss} and Eq.~\eqref{eq:Gauss2} with $V_0=100$ MeV, $F_0=3$ MeV, and $s=s_f=3$ fm. 
 The excitation energy $\epsilon$ is set to be 1 MeV. 
For the eigenvector continuation, we take $N_{\rm EC}=5$ and use $F_0$, randomly chosen in the range of 
(1 MeV, 5 MeV), while keeping the values of $V_0$, $s$, and $s_f$. 
That is, we vary the coupling strengths $F_0$ between the channels 1 and 2 while keeping the 
diagonal potential. 
We take 5 samples and take an ensemble average.
As in the single-channel case, one can see that the eigenvector continuation well reproduces the 
exact result, from well below the barrier to above the barrier. In particular, the exponential 
energy dependence of the penetrability at energies well below the barrier is successfully 
reproduced. 
The reduction in the computational time with the 
eigenvector continuation 
is by a factor of about 20, 
which is 
somewhat larger than the reduction factor in the single-channel case. 
\footnote{The reduction factor is about the factor 10 when the number 
of mesh points is 302. This value is similar to the one in the 
single-channel case.}
For the two-channel calculation considered here, 
the eigenvector continuation 
with $N_{\rm EC}=3$ already leads to a good reproduction 
of the exact result, which is further improved 
by increasing  $N_{\rm EC}$ to 5, as is shown in Fig. 4(c). 
The wave functions at $E=V_0$ are shown in Fig. 5. The left and the right panels 
show the wave functions for the channel 1, $\phi_1(x)$, and the channel 2, $\phi_2(x)$, respectively. 
As in the single-channel problem shown in Fig. 3, one can see that the eigenvector continuation 
well reproduces the wave functions, for both the channels. 

\section{Summary} \label{sec:Summary}

We have constructed an emulator for a scattering problem based on the eigenvector continuation 
and the Kohn variational principle 
in the discrete basis formalism. 
We have applied the emulator to a multi-channel scattering problem in one-dimension with a Gaussian barrier. 
To implement the eigenvector continuation, for a single-channel problem we have varied the barrier height. 
On the other hand, for a two-channel problem, we have varied the strength of the coupling potential while keeping 
the barrier height. We have demonstrated that the eigenvector continuation well reproduces the exact solutions 
both for the single-channel and the two-channel problems. In particular, it successfully reproduces the exponential energy 
dependence of the penetrability at energies well below the barrier. 

In this paper, we have considered a simple one-dimensional coupled-channels model. 
It would be straightforward to extend the emulator presented in this paper to more realistic scattering 
problems in three-dimension \cite{Hagino1999,Hagino2012,Hagino2022}. This will be useful e.g., in fitting 
experimental data to seek for optimal deformation parameters of colliding nuclei \cite{Gupta2020,Gupta2023}, 
which requires many calculations with different values of deformation parameters. This corresponds to 
varying coupling strengths while keeping the diagonal potential, as we have considered in Sec.~\ref{subsec:TwoChannEC} in this 
paper. We will report on this extension in a separate publication.

\begin{acknowledgments}
We thank G.F. Bertsch and X. Zhang  
for useful discussions. 
This work was supported in part 
by the RIKEN TRIP initiative (Nuclear Transmutation), 
and by 
JSPS KAKENHI Grant Numbers JP22K14030, JP23K03414, 23KJ1212, and 25H01511. 
S.Y. acknowledges JGC-Saneyoshi Scholarship Foundation.

\end{acknowledgments}

\bibliography{ref}
\end{document}